\documentclass[prd,twocolumn,groupedaddress,noshowpacs,nofootinbib]{revtex4}
\usepackage{graphicx}
\usepackage{dcolumn}
\usepackage{amssymb}
\usepackage{mathrsfs}
\usepackage{amsmath}
\usepackage{epsfig}
\usepackage[dvips]{color}
\usepackage{hhline}

\begin{document}

\title{	Peccei-Quinn symmetry with residual symmetries}

\author{Pei-Hong Gu}

\email{phgu@seu.edu.cn}

\affiliation{School of Physics, Jiulonghu Campus, Southeast University, Nanjing 211189, China}

\begin{abstract}

So far the null results from axion searches have enforced a huge hierarchy between the Peccei-Quinn and electroweak symmetry breaking scales. Then the inevitable Higgs portal poses a large fine tuning on the standard model Higgs scalar. Now we find if the Peccei-Quinn global symmetry has a set of residually discrete symmetries, these global and discrete symmetries can achieve a chain breaking at low scales such as the accessible TeV scale. This novel mechanism can accommodate some new phenomena including a sizable coupling of the standard model Higgs boson to the axion.

\end{abstract}


\maketitle

\section{ Introduction}

The most popular solution to the so-called strong CP problem is to introduce a global Peccei-Quinn  (PQ) symmetry \cite{pq1977}, which is traditionally called the $U(1)_{\textrm{PQ}}^{}$ symmetry. Through the color anomaly \cite{adler1969,bj1969,ab1969}, the Nambu-Goldstone (NG) boson from the spontaneous PQ symmetry breaking picks up a mass and hence becomes a pseudo NG boson, the axion \cite{weinberg1978,wilczek1978}. However, the axion has not been observed in any experiments. This stringently constrains the strength of the interactions between the axion and the standard model (SM) particles \cite{kg2010,pdg2020}. In fact the original PQ model with two Higgs doublets was shortly ruled out. Currently the Kim-Shifman-Vainstein-Zakharov (KSVZ) model \cite{kim1979,svz1980} and the Dine-Fischler-Srednicki-Zhitnitsky (DFSZ) model \cite{dfs1981,zhitnitsky1980} are widely adopted for decreasing the coupling of the axion to the SM.

In the KSVZ and DFSZ invisible axion models, the spontaneous PQ symmetry breaking should take place far far above the weak scale to evade all experimental limits from the mixing of the axion to the $\pi$ and $\eta$ mesons \cite{kg2010,pdg2020}. Accordingly, the accompanying new particles in these models will be very heavy, i.e. of the order of the PQ symmetry breaking scale, unless the related couplings are artificially chosen to be very small. Hence they will be completely isolated from experimental verification. On the theoretical point of view, it is also difficult to naturally explain the huge hierarchy between the PQ and electroweak symmetry breaking scales. With such hierarchy, unless the inevitable Higgs portal has an extremely small coupling, its contribution should have a large cancellation with the rarely quadratic term of the SM Higgs scalar. This problem plagues all of the invisible axion models. In this sense, we probably should say the high-scale PQ symmetry breaking provides a solution to the strong CP problem by paying a price of additionally fine tuning on the SM Higgs scalar.

In an early work \cite{mrs2001}, the authors tried to realize a theory with low-scale PQ symmetry breaking by introducing a large extra dimension and an anomalous $U(1)_A^{}$ gauge symmetry. In this model, two Higgs scalars in our brane could obtain their hierarchical vacuum expectation values (VEVs) after another Higgs scalar developed its VEV in a distant brane. The $U(1)_A^{}$ gauge field then could eat the most fraction of the NG boson coupling to a pair of new colored fermions. This could give a suppressed coupling of the axion to the new colored fermions and hence could allow an invisible axion from a low-scale PQ symmetry breaking. This model would be more attractive if it had no gauge anomalies.

In the present work, we shall propose a novel mechanism to lower the PQ symmetry breaking scale without any anomalous gauge symmetries. The key in our mechanism is to make the PQ global symmetry have a set of residually discrete symmetries by assigning proper PQ charges to the related Higgs scalars. Then the axion can appear if and only if the PQ symmetry and its residual symmetries are all broken spontaneously. In this chain breaking of the PQ and residual symmetries, the Higgs scalar(s) coupling to the colored fermions can only contribute a tiny fraction to the axion. As a result, the low breaking scales of these global and discrete symmetries can still allow a highly suppressed coupling of the axion to the colored fermions. In particular, when the PQ symmetry breaking scale is near the electroweak one, the SM Higgs boson can sizably couple to the axion through the Higgs portal, meanwhile, the new fermions in the KSVZ model or the new scalars in the DFSZ model can occur at the TeV scale to be verified experimentally.

\section{The chain breaking of a global symmetry and its residual symmetries}

For demonstration we first consider a simple system with two singlet scalars,  
\begin{eqnarray}
\label{global1}
\sigma_1^{}(1)\,,~~\sigma_n^{}(q_n^{})\,.
\end{eqnarray}
Here and thereafter the numbers in the brackets following the fields are the charges under a $U(1)_{G}^{}$ global symmetry. In order to realize this global symmetry, we do not constrain the model at renormalizable level. Instead, we give the scalar potential up to dimension-$(1+q_n^{})$, i.e.
\begin{eqnarray}
\label{potential1}
V&=& \mu_1^2 \sigma^\dagger_{1} \sigma_1^{} + \lambda_{11}^{}( \sigma^\dagger_{1}\sigma_1^{})^2_{}  + \mu_n^2 \sigma^\dagger_{n} \sigma_n^{}+ \lambda_{nn}^{} (\sigma^\dagger_{n}\sigma_n^{})^2_{} \nonumber\\
&&+ 2\lambda_{1n}^{} \sigma^\dagger_{1} \sigma_1^{} \sigma^\dagger_{n} \sigma^{}_n  + \frac{1}{\Lambda^{q_n^{}-3}_{}} (\sigma^\dagger_{n}\sigma^{q_n^{}}_{1}+\textrm{H.c.})\nonumber\\
&&+ \sum_{x+
y=3}^{\textrm{min}\{\frac{1}{2}q_n^{},\frac{1}{2}(q_n^{}+1)\}}\frac{c_{xy}^{}}{\Lambda^{2x+2y-4}_{}} (\sigma_1^\dagger \sigma_1^{})^x_{} (\sigma_n^\dagger \sigma_n^{})^y_{}  \,,
\end{eqnarray}
where the last but one term actually defines the global symmetry (\ref{global1}).

For a proper parameter choice, we can require the scalar $\sigma_n^{}$ to develop the VEV before the other scalar $\sigma_1^{}$. In this case, the initially global symmetry $U(1)_G^{}$ can have a residually discrete symmetry $Z_{q_n^{}}^{}$ due to the last but one term in Eq. (\ref{potential1}). The symmetry breaking pattern is nothing but 
\begin{eqnarray}
\label{breaking1}
U(1)_G^{}\stackrel{}{-\!\!\!-}\!\!\!\stackrel{\langle\sigma_n^{}\rangle}{-\!\!\!-}\!\!\!\stackrel{}{ \rightarrow} Z_{q_n^{}}^{}\stackrel{}{-\!\!\!-}\!\!\!\stackrel{\langle\sigma_1^{}\rangle}{-\!\!\!-}\!\!\!\stackrel{}{ \rightarrow} I \,.
\end{eqnarray}
After the scalars $\sigma_{1,n}^{}$ acquire their VEVs $v_{1,n}^{}$, they can be expressed by
\begin{eqnarray}
\label{vsp1}
\sigma_1^{}=\frac{1}{\sqrt{2}}(v_1^{}+S_1^{} + i P_1^{})\,,~~\sigma_n^{}=\frac{1}{\sqrt{2}}(v_n^{}+S_n^{} + i P_n^{})\,,
\end{eqnarray}
with $S_{1,n}^{}$ being the Higgs bosons while $P_{1,n}^{}$ being the pseudo scalars.

Since the symmetry breaking (\ref{breaking1}) only involves one global symmetry, the two pseudo scalars $P_{1,n}^{}$ should give one massless NG boson and one massive pseudo scalar. In fact, by taking the extreme condition into account, 
\begin{eqnarray}
\frac{\partial V }{\partial v_1^{}}=0\,,~~\frac{\partial V }{\partial v_n^{}}=0\,,
\end{eqnarray}
we can deduce the mass terms of the two pseudo scalars $P_{1,n}^{}$ to be 
\begin{eqnarray}
V &\supset & -\frac{v_1^{q_n^{}-2}}{4 (\sqrt{2}\Lambda)^{q_n^{}-3}_{} v_n^{}}\left(q_n^{} v_n^{} P_1^{} - v_1^{} P_n^{}\right)^2_{} \,.
\end{eqnarray}
Therefore, the mass eigenstates should be 
\begin{eqnarray}
\label{axion1}
\hat{P}_1^{}=\frac{q_n^{} v_n^{} P_1^{} - v_1^{} P_n^{}}{\sqrt{v_1^2 + q^2_{n} v_n^2 }} \,, ~~a=\frac{v_1^{} P_1^{} + q_n^{} v_n^{} P_n^{}}{\sqrt{v_1^2 + q^2_{n} v_n^2 }}\,,
\end{eqnarray}
with the mass eigenvalues, 
\begin{eqnarray}
 m_{\hat{P}_1^{}}^{2}= -\frac{v_1^{q_n^{}-2} (v_1^2 + q^2_{n} v_n^2 )}{2 (\sqrt{2} \Lambda)^{q_n^{}-3}_{} v_n^{}}\,,~~m_a^{}=0\,.
 \end{eqnarray}
Remarkably, $P_1^{}$ rather than $P_n^{}$ can contribute a dominant fraction to $\hat{P}_1^{}$ but a tiny fraction to $a$ if $v_{1,n}^{}$ and $q_n^{}$ match the condition $v_1^{}\ll  q_{n}^{} v_n^{}$. In this case, we can still have the flexibility to assign $q_n^{} \gg 1$ even if we take $v_1^{}\sim v_n^{}$.

In the above effective theory, the $U(1)_{G}^{}$ charge $q_n^{}$ is expected to be a very big number. If such number requires an explanation, it can be induced in some renormalizable models. For example, we can replace the scalar potential (\ref{potential1}) by the following one,
\begin{eqnarray}
\label{potential2}
V&=&\sum_{i,j=1}^{n} \left(\mu_i^2 \sigma_i^\dagger \sigma_i^{} + \lambda_{ij}^{} \sigma_i^\dagger \sigma_i^{} \sigma_j^\dagger \sigma_j^{} \right)\nonumber\\
&&+ \sum_{i=2}^{n}\rho_{i}^{}\left( \sigma_i^\dagger \sigma_{i-1}^{z_i^{}}+\textrm{H.c.}\right),~~(z_i^{}=2~\textrm{or}~3)\,. 
\end{eqnarray}
Here the singlet scalars $\sigma_{1,...,n}^{}$ carry the $U(1)_G^{}$ charges as below,
\begin{eqnarray}
\sigma_1^{}(1)\,,~~\sigma_{i}^{}(q_i^{}) ~~\textrm{with}~~q_i^{}=\prod_{j=2}^{i\geq 2 } z_j^{}\,.
\end{eqnarray}
Then we can require the symmetry breaking pattern to be 
\begin{eqnarray}
\label{breaking2}
U(1)_G^{}\stackrel{}{-\!\!\!-}\!\!\!\stackrel{\langle\sigma_n^{}\rangle}{-\!\!\!-}\!\!\!\stackrel{}{ \rightarrow} Z_{q_n^{}}^{}\stackrel{}{-\!\!\!-}\!\!\!\stackrel{\langle\sigma_{n-1}^{}\rangle}{-\!\!\!-\!\!\!-}\!\!\!\stackrel{}{ \rightarrow} Z_{q_{n-1}^{}}^{}\, ...\,  Z_{q_2^{}}^{} \stackrel{}{-\!\!\!-}\!\!\!\stackrel{\langle\sigma_1^{}\rangle}{-\!\!\!-}\!\!\!\stackrel{}{ \rightarrow} I \,.
\end{eqnarray}

By inserting the expressions 
\begin{eqnarray}
\label{vsp2}
\sigma_i^{}=\frac{1}{\sqrt{2}}(v_i^{}+S_i^{} + i P_i^{})\,,~~(i=1,...,n)\,,
\end{eqnarray}
into the scalar potential (\ref{potential2}), we in principle can solve the VEVs $v_{1,...,n}^{}$ by minimising the potential, i.e. $\partial V/\partial v_i^{}=0$, and then can obtain the mass terms of the Higgs bosons $S_{1,...,n}^{}$ and the pseudo scalars $P_{1,...,n}^{}$. According to the $U(1)_{G}^{}$ global symmetry breaking, the $n$ pseudo scalars $P_{1,...,n}^{}$ should leave $n-1$ massive pseudo scalars $\hat{P}_{1,...,n-1}^{}$ and one massless NG boson $a$. For simplicity we do not try to give the complicated expressions of $\hat{P}_{1,...,n-1}^{}$. Instead we only give the elegant expression of $a$, i.e. 
\begin{eqnarray}
\label{axion2}
a=\frac{v_1^{} P_1^{}+\sum_{i=2}^{n}q_i^{} v_i^{} P_i^{} }{\sqrt{v_1^2 + \sum_{i=2}^{n}(q^{}_{i} v_i^{})^2_{} }}~~\textrm{with}~~m_a^{}=0\,.
\end{eqnarray}
Clearly, $P_1^{}$ can only give a tiny contribution to $a$ if $q_n^{} \gg 1$ and $v_1^{}\sim v_2^{} \sim ... \sim v_{n-1}^{}\sim v_n^{}$.

We emphasize that the large $U(1)_G^{}$ charges in the above demonstrations will not plague any problems because the $U(1)_G^{}$ global symmetry has no perturbation requirement.

\section{The invisible axion from chain symmetry breaking}

In order to identify the above $U(1)_G^{}$ symmetry with a PQ symmetry, we introduce a pair of new color-triplet fermions $Q_{L,R}^{}$ with a hypercharge $-1/3$ or $+2/3$. These new fermions $Q_{L,R}^{}$ then can have a Yukawa coupling with the previously singlet scalar $\sigma_1^{}$, i.e. 
\begin{eqnarray}
\label{yukawa1}
\mathcal{L} \supset -y_Q^{} \left(\sigma_1^{} \bar{Q}_{L}^{} Q_{R}^{} + \textrm{H.c.}\right)~~\textrm{for}~~Q_{L}^{}(0)\,,~~Q_{R}^{}(-1)\,.
\end{eqnarray}
After the singlet scalar $\sigma_1^{}$ acquires its VEV $v_1^{}$, the chiral fermions $Q_{L,R}^{}$ can form a mass eigenstate, i.e.
\begin{eqnarray}
\label{mass1}
\mathcal{L}&\supset& - M_Q^{}\left( \bar{Q}_L^{} Q_R^{} + \textrm{H.c.} \right)= - M_Q^{} \bar{Q} Q \nonumber\\
[2mm]
&&\textrm{with}~~M_Q^{} = \frac{1}{\sqrt{2}}y_Q^{} v_1^{}\,.
\end{eqnarray}
At the same time, the pseudo scalar $P_1^{}$ from the singlet scalar $\sigma_1^{}$, see Eq. (\ref{vsp1}) or (\ref{vsp2}), can couple to the axial-vector current of new quark $Q$, i.e.
\begin{eqnarray}
\mathcal{L} &\supset& -i \frac{y_Q^{}  }{ \sqrt{2}} P_1^{} \bar{Q}\gamma_5^{} Q  =- \frac{P_1^{}}{2v_1^{}}\partial_\mu^{}\left( \bar{Q}\gamma_5^{} \gamma^\mu_{} Q\right)  \,.
\end{eqnarray}
Here we have made use of the Dirac equation.

Clearly, if there were no chain breaking (\ref{breaking1}) or (\ref{breaking2}), this $U(1)_G^{}$ global symmetry would be the usual PQ symmetry and this model would be just a KSVZ invisible axion model, where the pseudo scalar $P_1^{}$ fully gives the axion $a$, while the VEV $v_1^{}$ fully determines the axion decay constant $f_a^{}$. Now as shown in Eq. (\ref{axion1}) or (\ref{axion2}), the fraction of the axion $a$ in the pseudo scalar $P_1^{}$ is just
\begin{eqnarray}
P_1^{}=\frac{v_1^{} }{f_\sigma^{}}a+...~~\textrm{with}~~f_\sigma^{}=\left\{\begin{array}{ll}\sqrt{v_1^2 + q^2_{n} v_n^2 } \\
[1mm]
\textrm{or}\\
[1mm]
\sqrt{v_1^2 + \sum_{i=2}^{n}(q^{}_{i} v_i^{})^2_{} }\,.\end{array}\right.
\end{eqnarray}
Thus the true coupling between the axion $a$ and the new quark $Q$ should be  
\begin{eqnarray}
\label{agg1}
\mathcal{L}\supset   - \frac{a}{f_a^{}}\partial_\mu^{}\left( \bar{Q}\gamma_5^{} \gamma^\mu_{} Q\right)  ~~\textrm{with}~~f_a^{}= 2 f_\sigma^{}\,.
\end{eqnarray}

We now apply the chain breaking  (\ref{breaking1}) or (\ref{breaking2}) to the DFSZ model with two Higgs doublets, i.e.
\begin{eqnarray}
\label{potential3}
\mathcal{L} \supset  -y_d^{} \bar{q}_L^{} \phi_d^{} d_R^{} - y_u^{} \bar{q}_L^{} \tilde{\phi}_u^{} u_R^{} -\rho \sigma_1^{} \phi_u^\dagger \phi_d^{}+\textrm{H.c.}\,.
\end{eqnarray}
Here the SM quarks and the two Higgs doublets carry the $U(1)_G^{}$ charges as below,
\begin{eqnarray}
&&q_L^{}(0)=\left[\begin{array}{c}u_L^{}\\
[2mm]
d_L^{}\end{array}\right], ~~d_R^{}(-1/2),~~u_R^{}(-1/2),\nonumber\\
[2mm]
&&\phi_d^{}(-1/2)=\left[\begin{array}{c}\phi_{d}^{+}\\
[2mm]
\phi_d^{0}\end{array}\right],~~\phi_u^{}(+1/2)=\left[\begin{array}{c}\phi_{u}^{+}\\
[2mm]
\phi_{u}^{0}\end{array}\right].
\end{eqnarray}
When these Higgs doublets develop their VEVs to drive the $SU(2)_{L}^{} \times U(1)_{Y}^{} \rightarrow U(1)_{em}^{}$ electroweak symmetry breaking, they can be rewritten by 
\begin{eqnarray}
\phi_d^{}&=&\left[\begin{array}{c}
\phi_d^{+}\\
[2mm]
(v_d^{}+h_{d}^{}+iP_d^{})/\sqrt{2}
\end{array}\right],\nonumber\\
[2mm]
\phi_u^{}&=&\left[\begin{array}{c}\phi_{u}^{+}\\
[2mm]
(v_u^{}+h_u^{}+iP_u^{})/\sqrt{2}
\end{array}\right],
\end{eqnarray}
with $v_{d,u}^{}$ being the VEVs, $h_{d,u}^{}$ being the Higgs bosons and $P_{d,u}^{}$ being the pseudo scalars.

The doublets $\phi_{d,u}^{}$ and the singlets $\sigma_{1,n}^{}$ or $\sigma_{1,...,n}^{}$ should totally leave four NG bosons according to the electroweak gauge symmetry breaking and the PQ global symmetry breaking. As usual, the three NG bosons,
\begin{eqnarray}
&&G^{\pm}_{}= \frac{v_u^{}}{v_\phi^{}}\phi^{\pm}_{u}+ \frac{v_d^{}}{v_\phi^{}} \phi^{\pm}_{d}\,,~~G^{0}_{}= \frac{v_u^{}}{v_\phi^{}}P^{}_{u}+ \frac{v_d^{}}{v_\phi^{}} P^{}_{d}\nonumber\\
&&\textrm{with}~~v_\phi^{}=\sqrt{v_d^2+v_u^2}=246\,\textrm{GeV}\,,\end{eqnarray} 
should be eaten by the three SM gauge bosons $W^{\pm}_{}$ and $Z^0_{}$. As for the forth one, the axion $a$, it should be 
\begin{eqnarray}
\label{axion3}
a&=&\frac{(v_d^{} P_u^{} - v_u^{} P_d^{})v_d^{}v_u^{}/v^2_{\phi} +v_1^{} P_1^{}+q_n^{} v_n^{} P_n^{} }{\sqrt{v_d^2 v_u^2 /v^2_{\phi} + f_\sigma^2}}\nonumber\\
&&\textrm{with}~~f_\sigma^{}=\sqrt{v_1^2 + (q^{}_{n} v_n^{})^2_{} }\,,\nonumber\\
[1mm]
\textrm{or}&&\nonumber\\
[1mm]
a&=&\frac{(v_d^{} P_u^{} - v_u^{} P_d^{})v_d^{}v_u^{}/v^{2}_{\phi} +v_1^{} P_1^{}+\sum_{i=2}^{n}q_i^{} v_i^{} P_i^{} }{\sqrt{v_d^2 v_u^2 /v^2_{\phi} + f_\sigma^2}}\nonumber\\
&&\textrm{with}~~f_\sigma^{}=\sqrt{v_1^2 + \sum_{i=2}^{n}(q^{}_{i} v_i^{})^2_{} }\,.
\end{eqnarray}
Then the couplings of the axion to the SM quarks can be given by 
\begin{eqnarray}
\mathcal{L} &\supset &-i \frac{y_d^{}  }{ \sqrt{2}} P_d^{} \bar{d}\gamma_5^{} d   + i \frac{y_u^{}  }{ \sqrt{2}} P_u^{} \bar{u}\gamma_5^{} u \nonumber\\
&=& -  \frac{P_d^{}}{2v_d^{}}\partial_\mu^{}\left( \bar{d}\gamma_5^{} \gamma^\mu_{} d\right) + \frac{P_u^{}}{2v_u^{}} \partial_\mu^{}\left(\bar{u}\gamma_5^{} \gamma^\mu_{} u\right)\nonumber\\
&=&- \frac{a}{f_a^{}} \left(\frac{v_d^{2}}{v_\phi^{2}}\partial_\mu^{} \bar{d}\gamma_5^{} \gamma^\mu_{} d +\frac{v_u^{2}}{v_\phi^{2}} \partial_\mu^{}\bar{u}\gamma_5^{} \gamma^\mu_{} u\right)\nonumber\\
&&\textrm{with}~~f_a^{}=2\sqrt{v_d^2 v_u^2/v_\phi^2 + f_\sigma^2}\simeq 2f_\sigma^{}\gg v_\phi^{}\,.
 \end{eqnarray}

Currently, the null results from axion searches have put a low limit on the axion decay constant, i.e.  $f_a^{}\gtrsim 10^{9}_{}\,\textrm{GeV}$. This constraint can be satisfied in our mechanism even if the PQ symmetry and its residual symmetries are all broken at the TeV scale. For example, in the renormalizable models (\ref{potential2}), we can simplify 
\begin{eqnarray}
v_{1,...,n}^{}\equiv v_\sigma^{}\,,~~z_{1,...,n}^{}\equiv 3\,,
\end{eqnarray}
to produce
\begin{eqnarray}
f_a^{}&\simeq& 2 v_\sigma^{} \sqrt{\frac{9^{n}_{}-1}{8}} =3\times 10^{9}_{}\,\textrm{GeV} \nonumber\\
&&\textrm{for}~~v_\sigma^{}=10\,v_\phi^{}~~\textrm{and}~~n=14\,.
\end{eqnarray}

\section{ Other phenomena}

In the usual KSVZ and DFSZ invisible axion models, the new particles should be of the order of the PQ symmetry breaking scale and hence they could not leave any signals to be verified experimentally, unless we do some fine tuning on the related couplings. However, our models can allow the PQ symmetry and its residual symmetries to be spontaneously broken at some low scales such as the TeV scale. These low-scale symmetry breaking could naturally lead to some observable phenomena at colliders.

The SM Higgs doublet $\phi$, which is one of the linear combinations of the two Higgs doublets $\phi_{d,u}^{}$ in the DFSZ model, 
\begin{eqnarray}
\label{twohiggs}
\phi = \frac{v_d^{}}{v_\phi^{}} \phi_d^{} + \frac{v_u^{}}{v_\phi^{}} \phi_u^{}\,,~~\eta = \frac{v_u^{}}{v_\phi^{}} \phi_d^{} - \frac{v_d^{}}{v_\phi^{}} \phi_u^{}\,,
\end{eqnarray}
can have the quartic couplings with the singlet scalars $\sigma_{1,n}^{}$ or $\sigma_{1,...,n}^{}$, besides its rarely quadratic term, i.e.
\begin{eqnarray}
V &\supset & \mu_\phi^2 \phi^\dagger_{}\phi + \lambda_{\sigma_1^{}\phi }^{}  \sigma_1^\dagger \sigma_1^{}\phi^\dagger_{}\phi+  \lambda_{\sigma_n^{}\phi }^{}  \sigma_n^\dagger \sigma_n^{}\phi^\dagger_{}\phi\,,\nonumber\\
\textrm{or}&&\nonumber\\
V &\supset & \mu_\phi^2 \phi^\dagger_{}\phi + \sum_{i=1}^{n}\lambda_{\sigma_i^{}\phi }^{}  \sigma_i^\dagger \sigma_i^{}\phi^\dagger_{}\phi\,.
\end{eqnarray}
With these Higgs portal, the full quadratic term of the SM Higgs doublet should be 
\begin{eqnarray}
V& \supset & \left(\mu_\phi^2  + \frac{1}{2} \lambda_{\sigma_1^{}\phi }^{}  v_1^2  + \frac{1}{2} \lambda_{\sigma_n^{}\phi }^{}  v_n^2 \right)\phi^\dagger_{}\phi\,,\nonumber\\
\textrm{or}&&\nonumber\\
V &\supset & \left(\mu_\phi^2  + \frac{1}{2} \sum_{i=1}^{n}\lambda_{\sigma_i^{}\phi }^{}  v_i^2 \right)\phi^\dagger_{}\phi\,.
\end{eqnarray}  
Obviously, for a high-scale PQ symmetry breaking, the Higgs portal should be extremely weak in order to give the SM Higgs doublet, otherwise, there should be a large cancellation among the terms in the brackets. In fact, we should take $\lambda_{\sigma_i^{}\phi}^{}\lesssim 10^{-14}_{}$ for $v_i^{}\gtrsim 10^{9}_{}\,\textrm{GeV}$ and $\mu_\phi^2=\mathcal{O}((100\,\textrm{GeV})^2_{})$. Now the Higgs portal couplings need not be too small because the PQ symmetry and its residual symmetries are allowed to be broken near the weak scale. Then the Higgs boson $h$ from $\phi$ can have a sufficient mixing with the Higgs bosons $S_{1,n}^{}$ from
 $\sigma_{1,n}^{}$ or $S_{1,...,n}^{}$ from $\sigma_{1,...,n}^{}$. This mixing of order $\lambda_{\sigma_i^{}\phi}^{} v_\phi^{}/v_i^{}$ is small enough so that $h$ can behave almost exactly like the SM Higgs boson. However, the effective coupling of the SM Higgs boson to the axion, 
 \begin{eqnarray}
V \supset  \frac{1}{2}\lambda_{ah}^{} v_\phi^{} h a^2_{}\,,
\end{eqnarray}  
can be expected to significantly affect the invisible decay width of the SM Higgs boson as well as the scattering of axion off electron. Note the annihilating and scattering processes by this effective coupling can safely decouple above the QCD scale and hence will not change the other phenomena of the invisible axion.

The new quark $Q$ in the KSVZ model now is naturally allowed at the TeV scale and also mixes with the down-type quarks $d$ or the up-type quarks $u$. Through its single or pair production, the TeV-scale $Q$ can be verified at the LHC \cite{atlas2012,cms2018}. For example, the search is sensitive to both charged current and neutral current processes, $pp\rightarrow  Qq \rightarrow Wqq'$ and $pp \rightarrow Qq\rightarrow Zqq'$ with a leptonic decay of the vector gauge boson. Note the new quark $Q$ can always have a TeV-scale mass in the usual KSVZ model as long as the Yukawa coupling in Eq. (\ref{yukawa1}) is small enough \cite{dmt2014}. However, the TeV-scale $Q$ in our mechanism can contribute to the $h  \rightarrow \gamma\gamma$ and $h  \rightarrow gg$ channels through the unsuppressed Yukawa and Higgs portal interactions.

As for the DFSZ model in our mechanism, it can become a typical two Higgs doublet model with rich phenomena \cite{bhp1990} after the PQ symmetry and its residual symmetries are all broken at the TeV scale. Actually the non-SM Higgs doublet $\eta$ in Eq. (\ref{twohiggs}) could be at the TeV scale without fine tuning the Higgs portal. Such Higgs doublet has been studied in a lot of literatures.

\section{ Conclusion}

In this work, we have proposed a low-scale PQ symmetry breaking without introducing any anomalous gauge symmetries. Specifically, the PQ global symmetry should have a set of residually discrete symmetries. In order to generate the axion, the PQ symmetry and its residual symmetries should be all broken spontaneously. In this chain breaking of the PQ and residual symmetries, the Higgs scalar(s) coupling to the colored fermions can only have a negligible contribution to the axion. So, the low breaking scales of these global and discrete symmetries can be consistent with a highly suppressed coupling of the axion to the colored fermions. For such PQ symmetry with residual symmetries, the SM Higgs boson can sufficiently couple to the axion through the Higgs portal, meanwhile, the new fermions in the KSVZ model or the new scalars in the DFSZ model can have the masses at the TeV scale.

\textbf{Acknowledgement}: The author would like to thank Yu Gao, Xiao-Gang He, Tianjun Li, Ernest Ma, Rabi N. Mohapatra, Utpal Sarkar and Qiaoli Yang for useful communications. This work was supported in part by the Fundamental Research Funds for the Central Universities.

\end{document}